\documentclass[superscriptaddress,aps,journal=prl,preprint]{revtex4-2}
\usepackage{graphicx}
\usepackage{epstopdf}
\usepackage{bm,amsmath,amssymb} 
\usepackage{color, soul} 
\usepackage{dcolumn}   
\usepackage{multirow}
\usepackage{makecell}
\usepackage{threeparttable}
\usepackage{verbatim}

\newcommand{\eg}{{\em e.\,g.}}
\newcommand{\ie}{{\em i.\,e.}}

\newcommand{\PSTO}{Pb$_x$Sr$_{1-x}$TiO$_3$}
\newcommand{\HZO}{Hf$_x$Zr$_{1-x}$O$_2$}
\newcommand{\OP}{O$^{p}$}
\newcommand{\ONP}{O$^{np}$}
\newcommand{\VP}{V$_{\rm O}^{p}$}
\newcommand{\VNP}{V$_{\rm O}^{np}$}

\begin{document}
\title{Modular development of deep potential for complex solid solutions}

\author{Jing Wu}
\thanks{These two authors contributed equally}
\affiliation{Key Laboratory for Quantum Materials of Zhejiang Province, Department of Physics, School of Science and Research Center for Industries of the Future, Westlake University, Hangzhou, Zhejiang 310030, China}
\author{Jiyuan Yang}
\thanks{These two authors contributed equally}
\affiliation{Key Laboratory for Quantum Materials of Zhejiang Province, Department of Physics, School of Science and Research Center for Industries of the Future, Westlake University, Hangzhou, Zhejiang 310030, China}
\author{Liyang Ma}
\affiliation{Key Laboratory for Quantum Materials of Zhejiang Province, Department of Physics, School of Science and Research Center for Industries of the Future, Westlake University, Hangzhou, Zhejiang 310030, China}
\author{Linfeng Zhang}
\affiliation{DP Technology, Beijing 100080, China}
\affiliation{AI for Science Institute, Beijing 100080, China}
\author{Shi Liu}
\email{liushi@westlake.edu.cn}
\affiliation{Key Laboratory for Quantum Materials of Zhejiang Province, Department of Physics, School of Science and Research Center for Industries of the Future, Westlake University, Hangzhou, Zhejiang 310030, China}
\affiliation{Institute of Natural Sciences, Westlake Institute for Advanced Study, Hangzhou, Zhejiang 310024, China}


\begin{abstract}{ 
The multicomponent oxide solid solution is a versatile platform to tune the delicate balance between competing spin, charge, orbital, and lattice degrees of freedom for materials design and discovery. The development of compositionally complex oxides with superior functional properties has been largely empirical and serendipitous, in part due to the exceedingly complex chemistry and structure of solid solutions that span a range of length scales. The usage of classical molecular dynamics (MD), a powerful statistical method, in computer-aided materials design has not yet reached the same level of sophistication as that in computer-aided drug design because of the limited availability and accuracy of classical force fields for solids. Here, we introduce the strategy of ``modular development of deep potential" (ModDP) that enables a systematic development and improvement of deep neural network-based model potential, termed as deep potential, for complex solid solutions with minimum human intervention. The converged training database associated with an end-member material is treated as an independent module and is reused to train the deep potential of solid solutions via a concurrent learning procedure. We apply ModDP to obtain classical force fields of two technologically important solid solutions, \PSTO~and \HZO. For both materials systems, a single model potential is capable of predicting various properties of solid solutions including temperature-driven and composition-driven phase transitions over a wide range of compositions. In particular, the deep potential of \PSTO~reproduces 
a few known topological textures such as polar vortex lattice and electric dipole waves in PbTiO$_3$/SrTiO$_3$ superlattices, paving the way for MD investigations on the dynamics of topological structures in response to external stimuli. MD simulations of \HZO~reveal a substantial impact of composition variation on both the phase transition temperature and the nature of the high-temperature nonpolar phase.
}
\end{abstract}

\maketitle
\newpage

\section{Introduction}
Since Rahman's pioneer work of molecular dynamics (MD) simulation of liquid argon using a Lennard-Jones potential~\cite{Rahman64pA405}, MD as a statistical method has been employed to investigate and comprehend a broad range of systems ranging from solid and liquid crystals~\cite{Parrinello80p1196,Parrinello81p7182,Komolkin94p4103}, electrolytes~\cite{Bedrov19p7940,Kim11p8590}, to polymers~\cite{Grest86p3628} and colloids~\cite{Zaccarelli02p041402}, and to biological molecules such as proteins~\cite{Adcock06p1589,Isralewitz01p224} and DNA~\cite{Beveridge94p246}. The popularity of MD simulations relies upon its capability to reveal dynamics of atomic-level phenomena over large temporal and spacial scales with moderate computational cost, much more efficient than first-principles methods such as density functional theory (DFT). 

At the very heart of any MD simulation is the classical force field that approximates interatomic interactions. The well-established force fields such as CHARMM~\cite{Brooks83p187}, AMBER~\cite{Case05p1668}, and GROMOS~\cite{Scott99p3596} that are applicable to biologically important molecules (amino acids and nucleic acids) have enabled the facile deployment of MD for computer-aided drug discovery (CADD)~\cite{Durrant11p1,Borhani12p15}. In comparison, though MD simulations are wildly used to comprehend solid-state materials, the usage of MD in computer-aided materials discovery (CAMD) has yet reached the industry-level maturity as that in CADD, largely due to the difficulty of developing accurate and transferable force fields. For instance, MD simulations are routinely employed in the identification of lead compounds in drug design through virtual screening of extensive compound libraries. This enables researchers to narrow down candidates based on their interaction with the target protein, which is possible due to the availability of classical force fields capable of modeling a diverse range of compounds. A comparable practice is not yet available in CAMD.

Taking the technologically important $AB$O$_3$-type perovskite oxides as an example, the twelve-fold oxygen-coordinated $A$-site and the six-fold coordinated $B$-site can host nearly every element of the periodic table. At the same composition, the crystal structures of many perovskites can be modulated by thermodynamic variables like the electric field, temperature, and stress. 
Such chemical and structural versatility allows fine tuning of the energy scales of various interactions, \ie, spin-orbit interaction, crystal field, and electron-phonon coupling, offering a platform to create a rich spectrum of emergent phenomena~\cite{Ramesh19p257} including but not limited to ferroelectricity~\cite{Cohen92p136}, ferromagnetism~\cite{Tomioka95p3626,Zener51p403}, multiferroicity~\cite{Wang03p1719,Wang15p087601}, and superconductivity~\cite{Maeno94p532,Bednorz88p585}. On top of the intrinsic competitions of spin, charge, orbital, and lattice degrees of freedom, multiple component solid solutions afford additional complexity arsing from configuration and composition degrees of freedom~\cite{Grinberg02p909,Grinberg04p220101,Grinberg07p37603,Nishimatsu16p114714}.  

The delicate balance between interactions of different degrees of freedom, being the origin of the functional properties of perovskite oxides, nevertheness imposes a daunting challenge for the development of transferable force fields for this class of materials. Significant advancements have been made in the development of classical force fields for ferroelectrics and their solid solutions. The shell model has been widely used to simulate ferroelectrics~\cite{Mitchell93p1031,Sepliarsky05p107}, where each atom is represented by a positively charged core linked to a negatively charged shell through either a harmonic or anharmonic spring. By fitting to first-principles results, the shell model parameters for a few ferroelectric perovskites have been established, including PbTiO$_3$~\cite{Sepliarsky02p36}, BaTiO$_3$~\cite{Tinte99p9679, Vielma13p174108}, KNbO$_3$~\cite{Sepliarsky95p4044}, Ba(Ti,Sr)O$_3$~\cite{Tinte04p3495}, Pb(Zr,Ti)O$_3$ ~\cite{Gindele15p17784}, and (1-$x$)Pb(Mg$_{1/3}$Nb$_{2/3}$O$_3$)-$x$PbTiO$_3$~\cite{Sepliarsky11p435902}. For Pb(Zr, Ti)O$_3$ solid solutions, a single shell model can effectively predict the lattice parameters and polarization magnitudes of PbTiO$_3$, PbZrO$_3$, and Pb(Zr$_{0.5}$Ti$_{0.5}$)O$_3$~\cite{Gindele15p17784}. The bond valence (BV) model potential based on the bond valence theory~\cite{Brown73p266,Brown76p1957,Liu13p102202,Liu13p104102} serves as another useful tool for simulating ferroelectric solid solutions. The temperature-dependent structural properties of solid solutions with different $x$, such as Ba$_x$Sr$_{1-x}$TiO$_3$~\cite{Wexler19p174109} and Ba$_x$Ca$_{1-x}$ZrO$_3$~\cite{Zhang22p214204}, can be accurately predicted with a single set of potential parameters using the BV model. The ReaxFF potential utilizing distance-dependent bond-order functions is adept at describing the creation and breaking of chemical bonds~\cite{vanDuin01p9396, Senftle16p15011}. Specifically, the ReaxFF model of BaTiO$_3$ has been developed to investigate the dynamics of oxygen vacancies at elevated temperatures~\cite{Akbarian19p18240} and piezoelectric response of BaTiO$_3$ thin films~\cite{Kelley20p024407}.

Conventional force field development starts with a set of analytical functions, either constructed in an ``ad hoc" fashion or inspired by some physical/chemical principles, and the force field parameters are then fitted to reproduce known experimental results and/or quantum mechanical calculations. Despite achieving many notable successes, the process of developing a classical force field can be laborious and time-consuming: it usually requires months or even years of effort to obtain a force field for a single material. The critical issue is that a force field of a fixed mathematical construction and a limited number of parameters often lacks the capability to faithfully represent the intrinsically high-dimensional interatomic potential of complex materials where the subtle competitions between different energy terms have dictating impacts on the structure-property-performance relationship. Another equally if not more serious issue concerns the transferability of the force field. For example, in the shell model, a popular force field used extensively in atomistic simulations of oxides, the same element Ti acquires distinct shell charges in different ferroelectric systems, \ie, $-1.625$ in BaTiO$_3$~\cite{Vielma13p174108}, $-1.58$ in Ba$_x$Sr$_{1-x}$TiO$_3$~\cite{Sepliarsky05p107,Dimou22p094104}, $-5.158$ in PbTiO$_3$~\cite{Sepliarsky05p107}, $-6.8449$ in ($1-x$)PbMg$_{1/3}$Nb$_{2/3}$O$_3$-$x$PbTiO$_3$~\cite{Sepliarsky11p435902}, and $-5.047$ in PbZr$_{1-x}$Ti$_{x}$O$_3$~\cite{Gindele15p17784}. The change in the shell charge of Ti is physically justified as it reflects the change in the local chemical bonding environment characterized by the degree of charge transfer. However, this indicates the force field parameters developed for one material system can not be easily transferred to another system. The lack of a systematic approach to extend and improve force fields is not limited to perovskite oxides; it is the main hurdle for integrating MD with CAMD. 

In this work, we introduce the strategy of ``modular development of deep potential" (ModDP) to systematically develop, extend, and improve the force field for multicomponent solid solutions, using Pb$_x$Sr$_{1-x}$TiO$_3$ (PSTO) and Hf$_x$Zr$_{1-x}$O$_2$ (HZO) as examples. Deep potential (DP)~\cite{Zhang18p143001,Zhang18p4441} is a deep neural network-based model potential that has demonstrated excellent representability of complex and highly nonlinear potential energy surface for 
various material systems, including metals and alloys~\cite{wang21p025003}, two-dimensional materials~\cite{Wu21p174107}, ferroelectric oxides~\cite{He22p064104,Wu21p024108,Xie22p11839,Richard22p054066}, and halide perovskites~\cite{Yang22pe202100841,Wu22fluctuations,Tuo22p12445}. 
The essence of ModDP is to treat the training database associated with a parameterized DP model as the fundamental entity. The training database that yields an accurate DP model of a constituent solid is considered an independent and complete module, and is reused to construct the initial training database for solid solutions, in analogous to ``code reusing" in modular programming. The concurrent learning procedure, DP-GEN, is then adopted for  automatic, iterative, and efficient updates of the training database of solid solutions whose vast configurational and chemical space is sampled via MD simulations combined with Monte Carlo swapping processes.
Specifically, when developing the DP model of PSTO, we reuse a published database of SrTiO$_3$ (STO) that contains thousands of configurations without adding any new STO configurations in ModDP. Similarly, a database of HfO$_2$ is reused when training the DP model of HZO. The DP models of PSTO and HZO obtained via ModDP have DFT-level accuracy, as confirmed by a number of tests. Moreover, the DP model of PSTO reproduces a few experimentally observed real-space topological textures such as polar vortices and dipole waves in PbTiO$_3$/SrTiO$_3$ superlattices, despite the training database involving no superlattice configurations. The HZO DP model is further extended to acquire the capability of simulating oxygen deficient HfO$_{2-\delta}$.
This work demonstrates that ModDP enables straightforward development of accurate DP models for complex solid solutions with little human intervention, paving the way for the establishment of pseudopotential-like module-based force field library.

\section{COMPUTATIONAL METHODS}
\subsection{DP and DP-GEN}
Details of the DP model and DP-GEN have been extensively discussed in previous studies~\cite{Zhang18p143001,Zhang18p4441,Zhang20p107206}. Here, we only highlight several key points. The DP model uses deep neural networks (DNNs) to map the local environment of an atom to its energy ($E_i$), the sum of which gives the total energy ($E$), $E = \sum_i E_i$.
This approach is a manifestation of the embedded atom concept, where the many-body character of the interactions is reflected in the non-linear dependence of $E_i$ on the environment. The DNN effectively captures the analytical dependence of $E_i$ on the coordinates of the atoms in the local environment. This means that the functional form of DP is inherently many-body in nature and cannot be decomposed into two-, three-, and many-body energy terms. The DP-GEN scheme is an iterative procedure with each cycle comprising three steps,  exploration via MD, labeling via DFT, and training via deep learning. A notable feature of DP-GEN is the training of an ensemble of DP models with different initial values of DNN hyperparameters. For a newly MD-sampled configuration at the exploration step, the ensemble of DP models produce an ensemble of predictions, and the maximum standard deviation of atomic forces $\boldsymbol{F}_i$ is used to construct an error indicator for labeling,
\begin{equation}
\mathcal{E}=\max _i \sqrt{\left\langle\left\|\boldsymbol{F}_i-\left\langle\boldsymbol{F}_i\right\rangle\right\|^2\right\rangle}
\end{equation}
where $\langle\ldots\rangle$ denotes the average of predictions of all DP models. Only configurations with $\mathcal{E}_{\mathrm{lo}}<\mathcal{E}<\mathcal{E}_{\mathrm{hi}}$ are labeled for DFT calculations. The DP-GEN process stops when all sampled conﬁgurations satisfy $\mathcal{E}<\mathcal{E}_{\mathrm{lo}}$. Therefore, the value of $\mathcal{E}_{\mathrm{lo}}$ determines the target accuracy of the DP model.

\subsection{Modular development of deep potential}
For simple materials systems containing only two or three elements, the DP-GEN scheme is efficient, and the force field development is less impacted by the construction of the initial training database as DP-GEN can automatically and iteratively update the training database. However, \PSTO~is a challenging case for DP-GEN.
First, the two end members, PbTiO$_3$ (PTO) and SrTiO$_3$ (STO), are drastically different. At room temperatures, tetragonal PTO (space group $P4mm$) is a prototypical ferroelectric with large spontaneous polarization of 0.75 C/m$^2$~\cite{Gavrilyachenko70p1203} along the $c$-axis and thus a large $c/a$ ratio of 1.071~\cite{Mabud79p49}. It undergoes a ferroelectric tetragonal to paraelectric cubic (space group $Pm\bar{3}m$) phase transition characterized by a reducing displacement of the Ti (Pb) atom relative to the center of surrounding rigid O$_6$ (O$_{12}$) cage with increasing temperature, and the transition temperature ($T_c$) is 765~K~\cite{Shirane51p265}. 
In comparison, STO is a quantum paraelectric, and the low-temperature tetragonal phase (space group $I4/mcm$) has TiO$_6$ octahedral units rotating around the $c$-axis in an antiferrodistortive pattern, and the $c/a$ ratio is only 1.001~\cite{Cao00p387}. When the temperature is above 105~K~\cite{Riste71p1455}, STO adopts a cubic phase with TiO$_6$ octahedral tilt angle ($\phi$) being zero on average. The distinct structural properties of PTO and STO indicate these two compounds have markedly different potential energy surfaces, though both involving Ti-O bonds. Consequently, an initial database containing random configurations of PTO, STO, and \PSTO~leads to poor DP models after the training step in the first cycle of DP-GEN; MD simulations with those DP models tend to generate unphysical configurations. 

The ModDP protocol aims to minimize human intervention. As illustrated in Fig.~1, we first obtain the ``converged" training database for PTO and STO, respectively, using the standard DP-GEN scheme. The converged PTO database consists of 40-atom $2\times2\times2$ supercells of tetragonal and cubic phases. It is noted that for STO, we use a published database with 3538 configurations~\cite{He22p064104} including 40-atom $Pm\bar{3}m$ supercells and 20-atom $I4/mcm$ supercells, and the corresponding DP model has a very tight force convergence threshold of $\mathcal{E}_{\mathrm{lo}}=0.05$~eV/\AA. In contrast, using a moderate value of $\mathcal{E}_{\mathrm{lo}}=0.12$~eV/\AA~is enough to obtain a decent DP model of PTO after 20 iterations that well reproduce many thermodynamics properties and temperature-driven phase transitions in PTO; the final training database of PTO contains 13021 configurations. The initial database employed in DP-GEN to develop a single DP model applicable to \PSTO~of varying $x$ then contains the complete PTO and STO databases and some random configurations of \PSTO~with $x=0.25, 0.50, 0.75$. To reduce the computational cost, we set $\mathcal{E}_{\mathrm{lo}}=0.12$~eV/\AA. 
At the exploration step, MD simulations are carried out to sample new configurations of \PSTO~($x=0.25, 0.50, 0.75$) from 100 to 900~K in combined with Monte Carlo (MC) swaps of Pb and Sr atoms with different swapping probabilities  (100--900~K in the Metropolis criterion).
The ModDP protocol turns out to be quite robust: the DP-GEN reaches the convergence after 12 iterations. We perform an additional training with $\mathcal{E}_{\mathrm{lo}}=0.05$~eV/\AA~to recover the accuracy needed to describe pure STO. We make our final training database and hyperparameters available through a public repository, DP Library~\cite{DPLibrary}.

\subsection{DFT and MD simulations}
All DFT calculations are performed with the Vienna Ab initio Simulation (VASP) package~\cite{Kresse96p11169,Kresse96p15} and the projected augmented wave (PAW)
method~\cite{Blochi94p17953,Kresse99p1758} is employed. Perdew-Burke-Ernzerhof functional modified for solids (PBEsol)~\cite{Perdew08p136406} within the generalized gradient approximation is used as the exchange-correlation functional for PSTO whereas PBE is chosen for HZO. An energy cutoff of 800 (600) eV and a $k$-point spacing of 0.3 (0.5)~\AA$^{-1}$ are sufﬁcient to converge the energy and atomic forces of PSTO (HZO) configurations. 

The optimized DP model is used to study the temperature-driven phase transition by performing MD simulations in the isobaric-isothermal ($NPT$) ensemble. A 10$\times$10$\times$10 supercell containing 5000 atoms is used for PSTO while a supercell of 6144 atoms is used to model HZO systems. All $NPT$ MD simulations are performed using LAMMPS~\cite{Plimpton95p1}, with temperature controlled via the Nos\'e-Hoover thermostat and the pressure controlled by the Parrinello-Rahman barostat. The timestep for the
integration of the equation of motion is 2~fs for all MD simulations. The pressure is maintained at 1.0 bar when simulating the temperature-driven phase transition.  At a given temperature, the equilibrium run is 50 ps, followed by a production run of 50 ps that is sufficiently long to obtain converged structural parameters. Additionally, we compared the computational efficiency of the COMB
(charge-optimized many-body) potential of HfO$_2$~\cite{Shan10p125328} and the
DP method. In the speed benchmark test, a supercell containing 1500 atoms is used with two AMD EPYC 7513 32-core processors. The COMB force field has a speed of 0.0152 seconds per step when enabling charge equilibration (Qeq), and 0.0022 seconds per step without Qeq. In comparison, the DP model demonstrates a slower speed of 0.0648 seconds per step. It is worth noting that the DP model's speed improves to 0.018 seconds per step when using a single NVIDIA GeForce RTX 2080Ti GPU, and further accelerates to 0.005 seconds per step when employing an NVIDIA V100 Tensor Core GPU.

\section{RESULTS AND DISCUSSIONS}
\subsection{Ground-state structural properties}
The final training database has 19119 configurations in which only 2560 new configurations of \PSTO~are produced from the DP-GEN process of PSTO. This shows ModDP speeds up the reach to convergence. 
The DP model of PSTO achieves an excellent fitting to DFT energies and forces, as the mean absolute error is 0.825~meV/atom for energy and 0.037 eV/\AA~for atomic force, respectively. Table~\ref{ground_PSTO} reports the DFT and DP values of structural parameters for PTO, STO, and \PSTO~($x$=0.25, 0.5, 0.75) in different phases. It is evident that the PSTO DP model well reproduces DFT results.

The ModDP protocol results in three DP models, applicable to pure PTO, pure STO, and PSTO, respectively. Three models share the same DNN architecture comprising three layers and 240 nodes per layer. We first compare the fitting performance of PTO and PSTO DP models. As shown in Fig.~\ref{fit}a-b, for configurations in the training database used to converge the DP model of PTO, the PSTO DP model yields distributions of absolute error in energy and atomic force nearly identical to the PTO DP model. Similarly, the STO and PSTO DP models have comparable fitting performance for the training database used to obtain the STO DP model (Fig.~\ref{fit}c-d). These demonstrate the superior representability and flexibility of DP that a single DNN architecture is capable of describing complex and highly nonlinear energy functionals of drastically different materials systems. 
 
\subsection{Phonon spectrum and phase transition}
The DP model of PSTO reproduces the DFT phonon spectra of STO and PTO of different phases, as shown in Fig.~\ref{Phonon_PSTO}a-d. It is noted that we use the DFT optimized lattice constants to compute the phonon dispersion relationships instead of experimental lattice constants~\cite{Ghosez99p836}.
The DP phonon spectrum of tetragonal STO in the space group of $I4/mcm$ has unstable phonon modes at $\Gamma$ within the harmonic approximation (Fig.~\ref{Phonon_PSTO}a), consistent with previous DFT studies~\cite{Tadano19p404}. For cubic STO, the imaginary frequencies of unstable polar modes at $\Gamma$ and the instabilities at R predicted by DP agree with DFT values (Fig.~\ref{Phonon_PSTO}b). The high-frequency phonon modes computed with DP and DFT are nearly identical. In the case of PTO, the PSTO DP model correctly predicts a dynamically stable PTO in the tetragonal phase (space group $P4mm$) possessing no imaginary frequencies over the whole Brillouin zone (Fig.~\ref{Phonon_PSTO}c). The phonon dispersion of cubic PTO has several well known features, \ie, the instability in the lowest-frequency branch is not strongly localized around $\Gamma$ and the unstable R-M branch is 
rather flat~\cite{Ghosez99p836,Lebedev09p362}, both nicely captured by DP (Fig.~\ref{Phonon_PSTO}d). These calculations validate that a single DP model can accurately predict the second derivatives of energy with respect to atomic displacements (interatomic force constants) for both STO and PTO.

The temperature-driven phase transition in STO is investigated with $NPT$ simulations using the PSTO DP model. The temperature-dependent lattice constants and TiO$_6$ octahedral tilt angle ($\phi$) are presented in Fig.~\ref{Phonon_PSTO}e, revealing a tetragonal to cubic phase transition at $\approx$200~K that is consistent with previous MD simulations using a DP model of STO. Recently, it was shown that adding the nuclear quantum effects into MD simulation can reduce the theoretical $T_c$ to 160~K~\cite{Wu22p224102}, closer to experimental value of 105~K. Figure~\ref{Phonon_PSTO}d displays the temperature dependence of spontaneous polarization ($P_s$) and local atomic displacements of Pb and Ti ($D_{\rm Ti}$ and $D_{\rm Pb}$) determined with $NPT$ DPMD simulations. The magnitudes of $P_s$, $D_{\rm Ti}$, and $D_{\rm Pb}$ all decrease with increasing temperature, and PTO becomes paraelectric at $\approx$600~K. The predicted $T_c$ is lower than experimental value of $765$~K but improves over the value of $\approx 400$~K predicted by a bond-valence model~\cite{Liu13p104102} and is comparable with $T_c$ of a shell model~\cite{Gindele15p17784}.

\subsection{Phase diagram of \PSTO~solid solutions}
Though both PTO ans STO have been extensively studied, 
the temperature and concentration ($T$-$x$) phase diagram of \PSTO~is much less investigated, particularly for compositions near the STO end member. Recently, a thermodynamic potential of \PSTO~has been developed to map out the $T$-$x$ phase diagram for a single-domain sample, revealing a new ferroelectric $R3m$ phase for $0.025<x<0.18$ in the low-temperature region ($<130$~K)~\cite{Shirokov22p110395}. We note that the coefficients of the thermodynamic potential were mostly obtained by fitting to experimental data. Here, we construct the $T$-$x$ phase diagram with a fully {\em ab initio} model potential by performing $NPT$ DPMD simulations for various compositions: in the range of $x=0$--0.3 at intervals of 0.05 and in the range of $x=0.3$--1.0 at intervals of 0.1, with results compiled in Fig.~\ref{PSTO_Tx}a. Consistent with the phase diagram obtained with the phenomenological theory, the $T$-$x$ phase diagram obtained with DPMD reveal several critical points. Two critical points are for the phase transitions in the end members. There is a multicritical point at $x=$~0.25 and $T=$~170~K where tetragonal $P4mm$, cubic $Pm\bar{3}m$, and rhombohedral $R3m$ phases converge. The $R3m$ phase has the polarization aligned along the [111] direction. Another multicritical point occurs at $x=$~0.15 and $T=$~120~K that separates tetragonal $I4/mcm$, cubic $Pm\bar{3}m$, and ferroelectric $R3m$ phases.  

Figure.~\ref{PSTO_Tx}b presents the evolution of [111] local atomic displacements ($D_{\rm Pb}$, $D_{\rm Sr}$, and $D_{\rm Ti}$) with increasing temperature in Pb$_{0.2}$Sr$_{0.8}$TiO$_3$, revealing a transition from ferroelectric $R3m$ to paraelectric $Pm\bar{3}m$ at $\approx$150~K that is close to the second order. Additionally, Pb atoms are most displaced, followed by Ti and Sr atoms. At a higher PTO concentration of $x=0.5$, the ferroelectric phase becomes tetragonal, and the temperature-driven transition from $P4mm$ to $Pm\bar{3}m$ as characterized by [001] atomic displacements becomes sharper, indicating a first-order-like phase transition (Fig.~\ref{PSTO_Tx}c). In the PTO-rich region, the PSTO DP model correctly predicts a reducing $T_c$ with decreasing $x$. As shown in the composition-driven phase transition (Fig.~\ref{PSTO_Tx}d), $x>0.5$ is required to obtain ferroelectric \PSTO~solid solutions at room temperatures. 

\subsection{PTO/STO superlattices}
The PTO/STO superlattices has become a model system to investigate real-space topological textures such as flux closures~\cite{Tang15p547}, vortices~\cite{Yadav16p198}, skyrmions~\cite{Das19p368}, and merons~\cite{Wang20p881}. These novel structural topologies can host various emergent phenomena such as chirality~\cite{Shafer18p915,Behera22peabj8030,Mccarter22p247601} and negative capacitance~\cite{Zubko16p524,Das21p194,Iniguez19p243} that may enable low-power electronics. Second-principle methods based on effective Hamiltonian and phase-field modeling using parameterized thermodynamic potential are the main theoretical tools to comprehend the ground-state properties of mesoscale topological structures in ferroic systems~\cite{Han22p63,Yadav16p198,Das19p368,Hong17p2246,Wang20p881,Das21p194}. However, it remains challenging to study the dynamics of topological domains in response to external stimuli with sufficiently high spacial/time resolution. The PSTO DP model developed in this work enables large-scale MD simulations of PTO/STO superlattices and successfully reproduce several topological textures.  

A 40$\times$20$\times$20 supercell consisting of 80,000 atoms is used to model a (PTO)$_{10}$/(STO)$_{10}$ superlattice (Fig.~\ref{Topo_struc}a). At a strain state that has averaged in-plane lattice constant $a_{\rm IP}$ of 3.937~\AA ~and $b_{\rm IP}$ of 3.930~\AA, the equilibrium configuration at 300~K obtained with DPMD simulations adopts an ordered polar vortex lattice with alternating vortex and antivortex (Fig.~\ref{Topo_struc}b), resembling the experimental observations~\cite{Yadav16p198}. After increasing the in-plane strain to $a_{\rm IP}=$~3.949~\AA, we observe a shift of vortex cores toward the PTO/STO interfaces (Fig.~\ref{Topo_struc}c). Finally, at a large tensile in-plane strain ($a_{\rm IP}=$~3.954~\AA), periodic electric dipole waves characterized by head-to-tail connected electric dipoles in the form of sine function emerge in the (PTO)$_{10}$/(STO)$_{10}$ superlattice (Fig.~\ref{Topo_struc}d). This agrees with a recent experiment where the scandate substrates (\eg, DyScO$_3$, SmScO$_3$, and NdScO$_3$) were employed to impose tensile epitaxial strains to realize electric dipole waves in PTO/STO superlattices~\cite{Gong21peabg5503}. We emphasize that the final training database does not contain any superlattice configurations. The ability of the PSTO DP model to predict strain-driven topological transition in PTO/STO superlattices serves as a strong evidence corroborating the accuracy and transferability of the force field. 

\subsection{DP model of Hf$_x$Zr$_{1-x}$O$_2$}
The discovery of robust nanoscale ferroelectricity in HfO$_2$-based~\cite{Boscke11p102903} and ZrO$_2$-based~\cite{Huang21p116536} thin films have revitalized the development for ferroelectric-based nanoelectronics owing to the excellent compatibility of HfO$_2$ and ZrO$_2$ with the modern complementary metal oxide semiconductor (CMOS) technology~\cite{Luo20p1391,Kim21peabe1341}. Due to the lanthanide contraction effect, ZrO$_2$ and HfO$_2$ have similar structural and chemical properties, and they can form single-phase solid solutions over the entire composition range. Both ZrO$_2$ and HfO$_2$ can adopt the polar $Pca2_1$ phase that is widely viewed as the ferroelectric phase in thin films~\cite{Park15p1811,Huan14p064111,Huang21p116536}. It was found that the mixed system of Hf$_x$Zr$_{1-x}$O$_2$ (HZO) can support ferroelectricity for a wide range of values of $x$~\cite{Mller12p4318,Park17p9973}. Moreover, Hf$_x$Zr$_{1-x}$O$_2$ thin films could be crystallized at lower temperatures than HfO$_2$-based thin films, beneficial for the integration process~\cite{Mller12p4318}. 

To demonstrate the robustness of ModDP, we further develop a DP model capable of describing HZO solid solutions. The DP-GEN scheme is used to converge the database of ZrO$_2$ that contains 9085 configurations of 96-atom supercells that cover $P2_1/c$, $Pbca$, $Pca2_1$, and $P4_2/nmc$ phases.
The initial database to develop HZO DP model includes the database of ZrO$_2$, a published database of HfO$_2$ (21768 configurations)~\cite{Wu21p024108}, and a number of random configurations of \HZO~with $x=0.25, 0.5, 0.75$.
We set $\mathcal{E}_{\mathrm{lo}}=0.15$~eV/\AA, and the exploration step in DP-GEN is carried out by running MD simulations augmented by MC swapping for \HZO~with $x=0.25, 0.5, 0.75$ at temperatures between 100--2800~K and pressures 0--20~GPa to sample the configuration space of HZO. The DP-GEN process results in 25809 new HZO configurations which also cover $P2_1/c$, $Pbca$, $Pca2_1$, and $P4_2/nmc$ phases. 

The accuracy of the DP model of HZO is confirmed by several tests. The DFT phonon spectra of HfO$_2$ and ZrO$_2$ in the ferroelectric $Pca2_1$ phase are reproduced by the HZO DP model (Fig.~\ref{Phonon_HZO}). The $Pca2_1$ phase has two symmetry-inequivalent oxygen atoms (see schematics in Fig.~\ref{Phonon_HZO}), three-fold coordinated polar oxygen (\OP) and fourfold-coordinated nonpolar oxygen (\ONP). Previous studies suggested two possible polarization switching mechanisms: the shift-inside (SI) pathway that has oxygen atoms moving inside two Hf atomic planes (Fig.~\ref{HfO}a) and the shift-across (SA) pathway that has polar oxygen atoms moving across the Hf atomic plane (Fig.~\ref{HfO}b)~\cite{Wei22p154101,Choe21p8}. As illustrated in Fig.~\ref{HfO}, the DP values of switching barriers for SI and SA mechanisms in HfO$_2$, Hf$_{0.5}$Zr$_{0.5}$O$_2$, and ZrO$_2$ are all in good agreement with DFT values. 


\subsection{DP model of HfO$_{2-\delta}$}
Defects such as oxygen vacancy can strongly impact the 
structural polymorphism kinetics and polarization switching dynamics in HfO$_2$-based thin films~\cite{Ma22p09374}. It is desirable to make the DP model of HZO applicable to oxygen-deficient HfO$_{2-\delta}$. In the same spirit of ModDP, a database of HfO$_{2-\delta}$ configurations is constructed by randomly removing oxygen atoms from supercells of HfO$_2$ in $P2_1/c$, $Pbca$, $Pca2_1$, and $P4_2/nmc$ phases as well as some intermediate configurations 
along the diffusion pathways of \VP~(vacancy at \OP~site) and \VNP (vacancy at \ONP~site). This HfO$_{2-\delta}$ database is then added to the HZO database, followed by a one-shot deep training. The resulted DP model demonstrates remarkable accuracy in predicting diffusion barriers of \VP~and \VNP~for multiple pathways (Fig.~\ref{defect-NBE}). We expect that utilizing the ModDP protocol can further extend the HZO DP model to simulate Hf$_x$Zr$_{1-x}$O$_{2-\delta}$.

Finally, we report in Fig.~\ref{phase-transition}a the temperature-driven phase transitions of HfO$_{2-\delta}$ and \HZO~($x=0.0, 0.25, 0.5, 0.75, 0.9, 1.0$) starting in the ferroelectric $Pca2_1$ phase, all obtained from DPMD simulations of 6144-atom supercells using the same DP model. The magnitude of the spontaneous polarization is gauged by the ensemble averaged displacement of polar oxygen ($\left<D({\rm O}^p) \right>$ with $\left<...\right>$ denoting the ensemble average over all \OP~atoms) relative to the center of the surrounding Hf$_4$ tetrahedron (see schematic in Fig.~\ref{phase-transition}b).
It is found that all materials systems possess a phase transition with increasing temperature, and HfO$_2$ and ZrO$_2$ have $T_c$ of approximately 900 K at a pressure of 1 bar.
We note that at a higher pressure of 5 GPa, the $T_c$ value of HfO$_2$ increases to $\approx$1900~K~\cite{Wu21p024108}, consistent with the predicted electric auxetic effect in HfO$_2$ where a hydrostatic pressure will promote the polarization~\cite{Liu20p197601}.
Importantly, our MD simulations reveal two types of phase transitions, $Pca2_1 \rightarrow P4_2/nmc$ in \HZO~with $x=0.0, 0.25$, and $Pca2_1 \rightarrow Pbcn$ in \HZO~with $x=0.5, 0.75, 0.9, 1.0$ (Fig.~\ref{phase-transition}b) where the $Pbcn$ phase is characterized by neighboring antiparallel \OP~atoms. Though $Pca2_1 \rightarrow P4_2/nmc$ is commonly viewed as the process responsible for the temperature-driven polar-nonpolar phase transition in ferroelectric hafnia and zirconia thin films, the $Pbcn$ phase was long postulated as an intermediate orthorhombic phase bridging $P4_2/nmc$ and $Pca2_1$ based on group-subgroup arguments~\cite{Trolliard11p264}. Moreover, as presented in Table~\ref{ground_HZO}, both DFT and DP predict that the $Pbcn$ phase has a lower energy than $Pca2_1$ in HfO$_2$. Interestingly, the $Pbcn$ phase becomes unstable relative to $Pca2_1$ and $P4_2/nmc$ in ZrO$_2$. These results explain why ZrO$_2$ undergoes $Pca2_1\rightarrow P4_2/nmc$ while HfO$_2$ has $Pca2_1\rightarrow Pbcn$ with increasing temperature. In comparison, Hf$_{0.5}$Zr$_{0.5}$O$_2$ has $Pbcn$ higher in energy than $Pca2_1$ but lower than $P4_2/nmc$ (Table~\ref{ground_HZO}), corroborating the temperature-driven $Pca2_1 \rightarrow Pbcn$ transition revealed from DPMD simulations. To fully resolve the details of phase transition mechanisms in hafnia and zirconia demand further experimental and theoretical investigations beyond the scope of current work. Nevertheless, the developed DP model can serve as a useful tool to provide atomic-level insight. Additionally, HfO$_{2-\delta}$ with an oxygen vacancy concentration of 3.125\% has $T_c$ substantially suppressed than that in pure HfO$_2$, and the value of $T_c$ is 800~K, comparable with the experimental value of $\approx$723 K in Y-doped HfO$_2$~\cite{Shimizu16p32931}. This hints at the importance of oxygen vacancy on the thermal stability of HfO$_2$-based ferroelectrics. 

\section{CONCLUSION}
Compositionally complex oxides and the designed hetrostructures containing complex oxides host a rich spectrum of functional properties that may enable new devices and technologies. To facilitate the discovery and optimization of multicomponent solid solutions with tailored physical properties to meet the requirements of applications, it is important to have easily deployable computational tools to probe the chemistry and structural complexity that often spans multiple scales to gain understanding of the structure-property relationship. We propose a ModDP protocol that allows for systematic and facile development of accurate DP models of complex solid solutions for large-scale MD simulations. The converged training database associated with a parameterized DP model of an end-member material is considered as the fundamental entity in ModDP, and is reused to construct the initial training database that speeds up
the reach to convergence of DP models of solid solutions without modifying the deep neural network architecture. The robustness of ModDP is demonstrated by the developed accurate force fields of \PSTO~and \HZO~where a single DP model is capable of describing various physical properties of solid solutions over a wide composition range. In particular, the DP model of PSTO enables future MD studies of the dynamics of topological textures in PTO/STO superlattices with atomic resolution, while the HZO DP model is a useful tool to probe the effects of composition on the structural polymorphism and temperature-driven phase transitions. When combined with high-throughput computations, we envision the ModDP protocol makes it possible to construct pseudopotential-like module-based force field library.

\begin{acknowledgments}
This work is supported by National Natural Science Foundation of China (12074319), National Key R\&D Program of China (2021YFA1202100), Natural Science Foundation of Zhejiang Province (2022XHSJJ006), and Westlake Education Foundation. The computational resource is provided by Westlake HPC Center.
\end{acknowledgments}




\bibliography{SL}


\clearpage
\newpage
\renewcommand{\arraystretch}{0.8}
\begin{table}[]
\centering
\caption{Lattice parameters ($a$ and $c$ in \AA) and energy ($E$ in eV/atom) of ground-state structures of STO, PTO, and \PSTO~in different phases computed with DFT and DP model of PSTO. For each $x$, two configurations of different cation arrangements of \PSTO~solid solutions are studied.
Absolute error in percentage.}
\label{ground_PSTO}
\tabcolsep=1cm
\resizebox{!}{0.6 \textwidth}{
\begin{tabular}{c|cccc}
\hline
\hline
STO        & Method     & $a$  & $c$     & $E$    \\
\hline
\multirow{3}{*}{$P m \bar{3} m$}  &DFT & 3.896   &    & -8.406527  \\
                    &DP          & 3.897  &    & -8.406596         \\
                    &Error       & 0.026\% &   &  0.0008\%  \\ 

\hline
\multirow{2}{*}{$I$4/$mcm$} &DFT & 3.887 & 3.907    & -8.407662 \\
                      &DP    & 3.888 & 3.909   &-8.407530     \\
                      &Error & 0.026\%  & 0.051\%   &  0.0016\%  \\
\hline
PTO  &Method  & $a$   & $c$  & $E$    \\
\hline
\multirow{3}{*}{$P m \bar{3} m$} &DFT & 3.922  &   & -7.951739  \\
                    &DP  & 3.924 &  &-7.952134  \\
                    &Error & 0.051\%  &    &  0.0050\%  \\
\hline
\multirow{3}{*}{$P$4$mm$} &DFT & 3.874    & 4.200   & -7.967686 \\
                &DP  & 3.874  & 4.199  &-7.964719 \\
                &Error & 0.0\%  & 0.024\%   &  0.0372\%  \\
\hline
Pb$_{0.25}$Sr$_{0.75}$TiO$_3$  &Method  & $a$  & $c$ & $E$    \\
\hline
\multirow{3}{*}{Configuration 1} &DFT & 3.911  &   & -8.294573 \\
                           &DP  & 3.915 &  &-8.295037  \\
                           &Error & 0.102\%  &  &  0.0056\%  \\
\hline
\multirow{3}{*}{Configuration 2} &DFT & 3.910 &   & -8.294127 \\
                                 &DP  & 3.913  &   &-8.294780 \\
                                 &Error & 0.077\%  &  &  0.0079\%  \\
\hline
Pb$_{0.5}$Sr$_{0.5}$TiO$_3$  &Method  & $a$  & $c$  & $E$\\
\hline
\multirow{3}{*}{Configuration 1} &DFT & 3.896  & 3.984  & -8.183353 \\
                                &DP  & 3.903 & 3.972  &-8.183974  \\
                          &Error & 0.180\%  &0.301\%  &0.0076\%  \\
\hline
\multirow{3}{*}{Configuration 2} &DFT & 3.895 & 3.998  & -8.184786 \\
                &DP  & 3.905  & 3.975  &-8.185180 \\
                &Error & 0.257\%  &0.575\%  &0.0048\%  \\
\hline
Pb$_{0.75}$Sr$_{0.25}$TiO$_3$  &Method  & $a$  & $c$  & $E$   \\
\hline
\multirow{3}{*}{Configuration 1} &DFT & 3.891  & 4.067  & -8.074845 \\
                           &DP  & 3.902 & 4.027  & -8.074593  \\
                           &Error & 0.283\%  &0.984\%  &0.0031\%  \\
\hline
\multirow{3}{*}{Configuration 2} &DFT & 3.891 & 4.061  & -8.074357 \\
                &DP  & 3.901  & 4.026  &-8.074181 \\
                &Error & 0.257\%  &0.862\%  &0.0022\%  \\
\hline
\hline
\end{tabular}}
\end{table}

\clearpage
\newpage
\renewcommand{\arraystretch}{0.8}
\begin{table}[]
\centering
\caption{Energy in meV/atom of $Pbcn$ and $P4_2/nmc$ relative to $Pca2_1$ in HfO$_2$, Hf$_{0.5}$Zr$_{0.5}$O$_2$, and ZrO$_2$ computed with DFT and DP using a 12-atom unit cell.}
\label{ground_HZO}
\tabcolsep=1cm
\begin{tabular}{cccc}
\hline
\hline
  & Method &$Pbcn$ &$P4_2/nmc$  \\
\hline
\multirow{2}{*}{HfO$_2$}  &DFT  &$-1.525$ &27.442  \\
                          &DP   &$-2.399$  &29.859  \\
\hline
\multirow{2}{*}{Hf$_{0.5}$Zr$_{0.5}$O$_2$}  &DFT  &9.155 &20.775 \\
                          &DP   &13.142  &14.384  \\
\hline
\multirow{2}{*}{ZrO$_2$}  &DFT  &21.300  &12.792  \\
                          &DP   &24.839  &11.839  \\
\hline
\hline
\end{tabular}
\end{table}

\clearpage
\newpage
\begin{figure}[]
\centering
\includegraphics[width=0.8\textwidth]{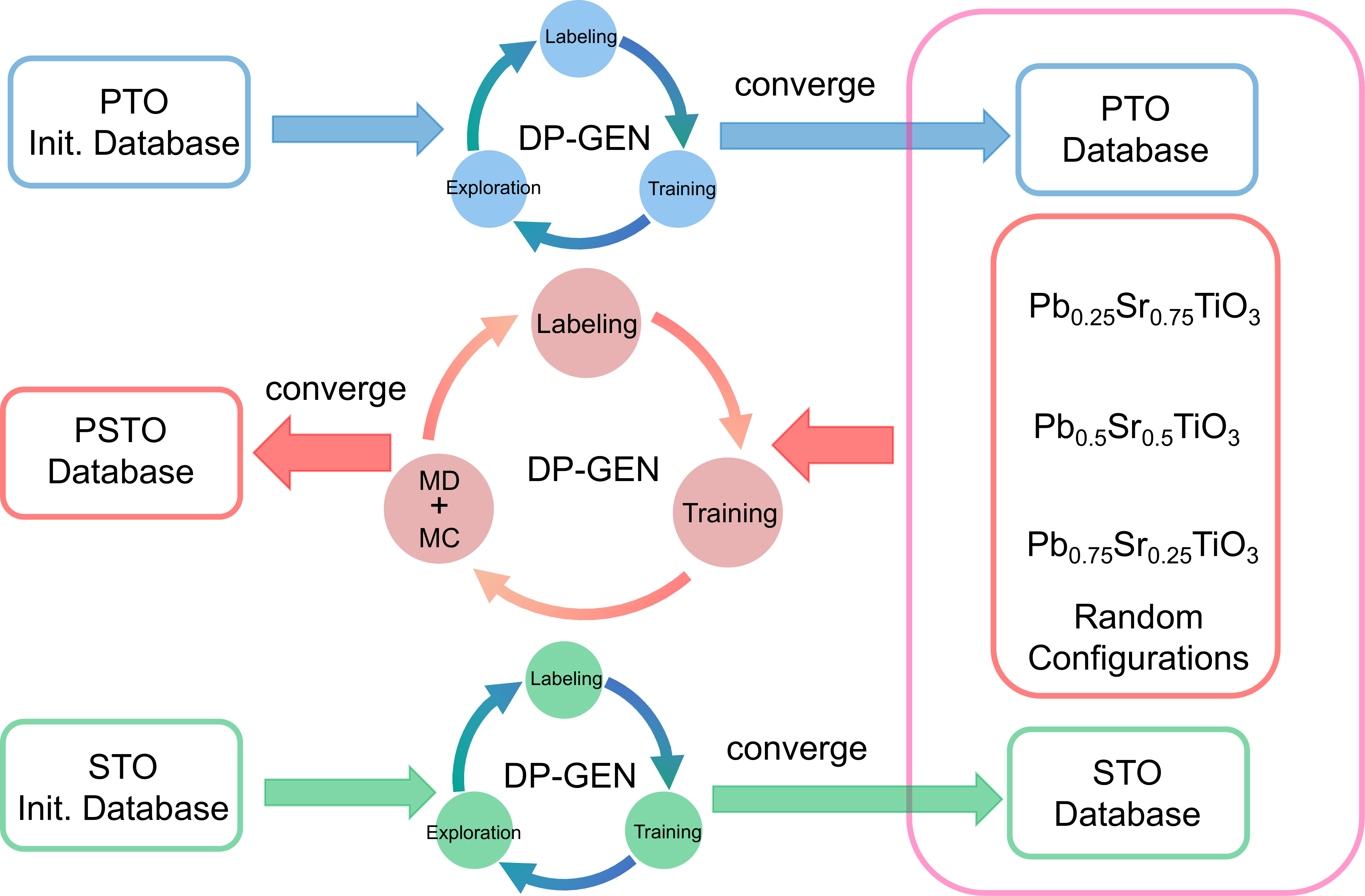}
\caption{Schematic of ModDP. The DP models of PTO and STO are first obtained with the DP-GEN scheme, respectively. The converged databases of PTO and STO are used to construct the initial training database in combination with some random configurations of \PSTO~($x=0.25, 0.50, 0.75$) to start the DP-GEN process to converge the DP model of PSTO.}
\label{Protocol_exploring}
\end{figure}

\clearpage
\newpage
\begin{figure}[]
\centering
\includegraphics[width=0.8\textwidth]{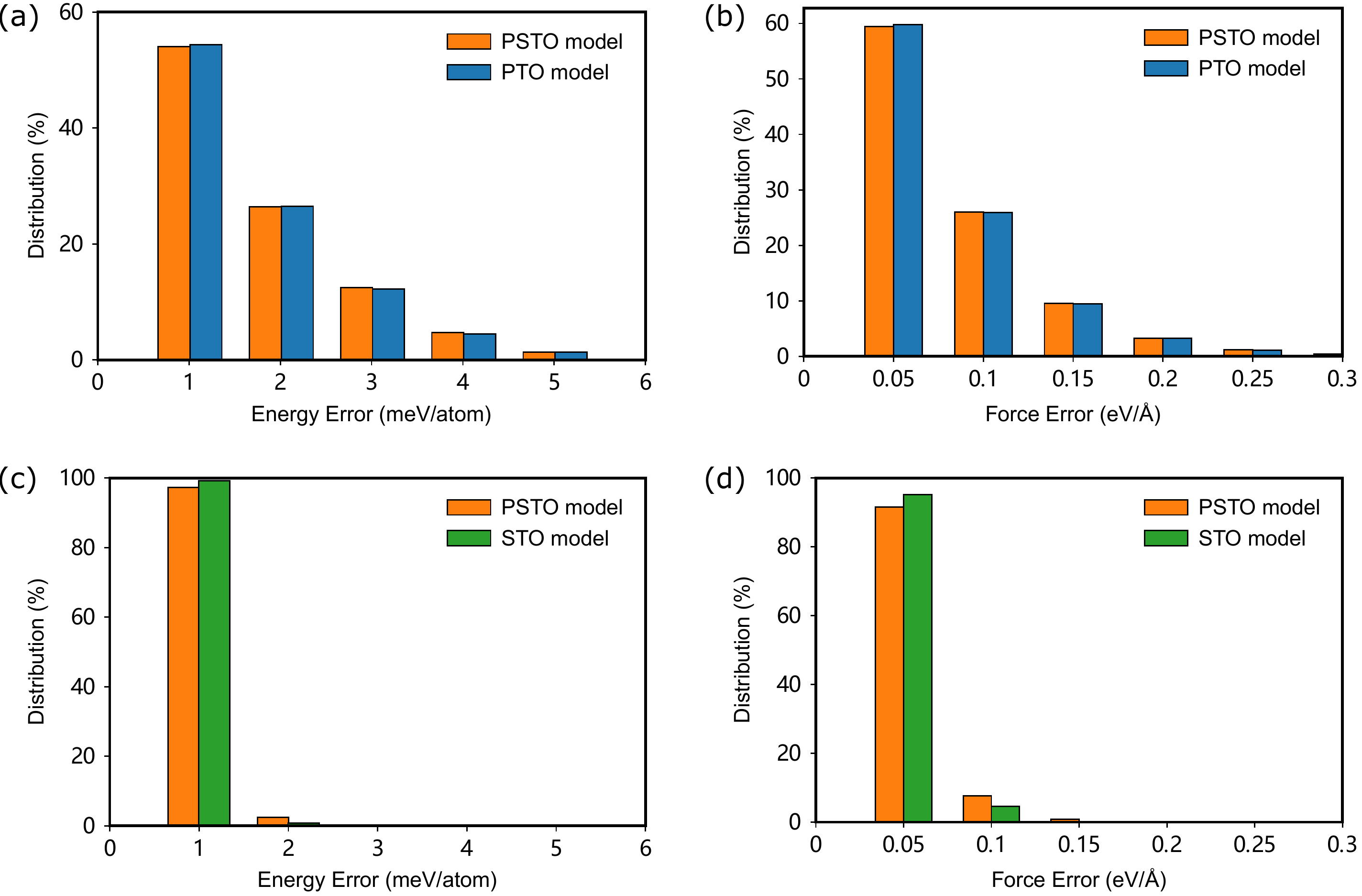}
\caption{Comparison of the distributions of the absolute error in (a) energy and (b) atomic force computed with PTO and PSTO DP models for configurations in the converged training database used to obtain the PTO DP model. Comparison of the fitting performance of STO and PSTO DP models in (c) energy and (d) atomic force for all STO configurations.
}
\label{fit}
\end{figure}

\clearpage
\newpage
\begin{figure}[]
\centering
\includegraphics[width=0.8\textwidth]{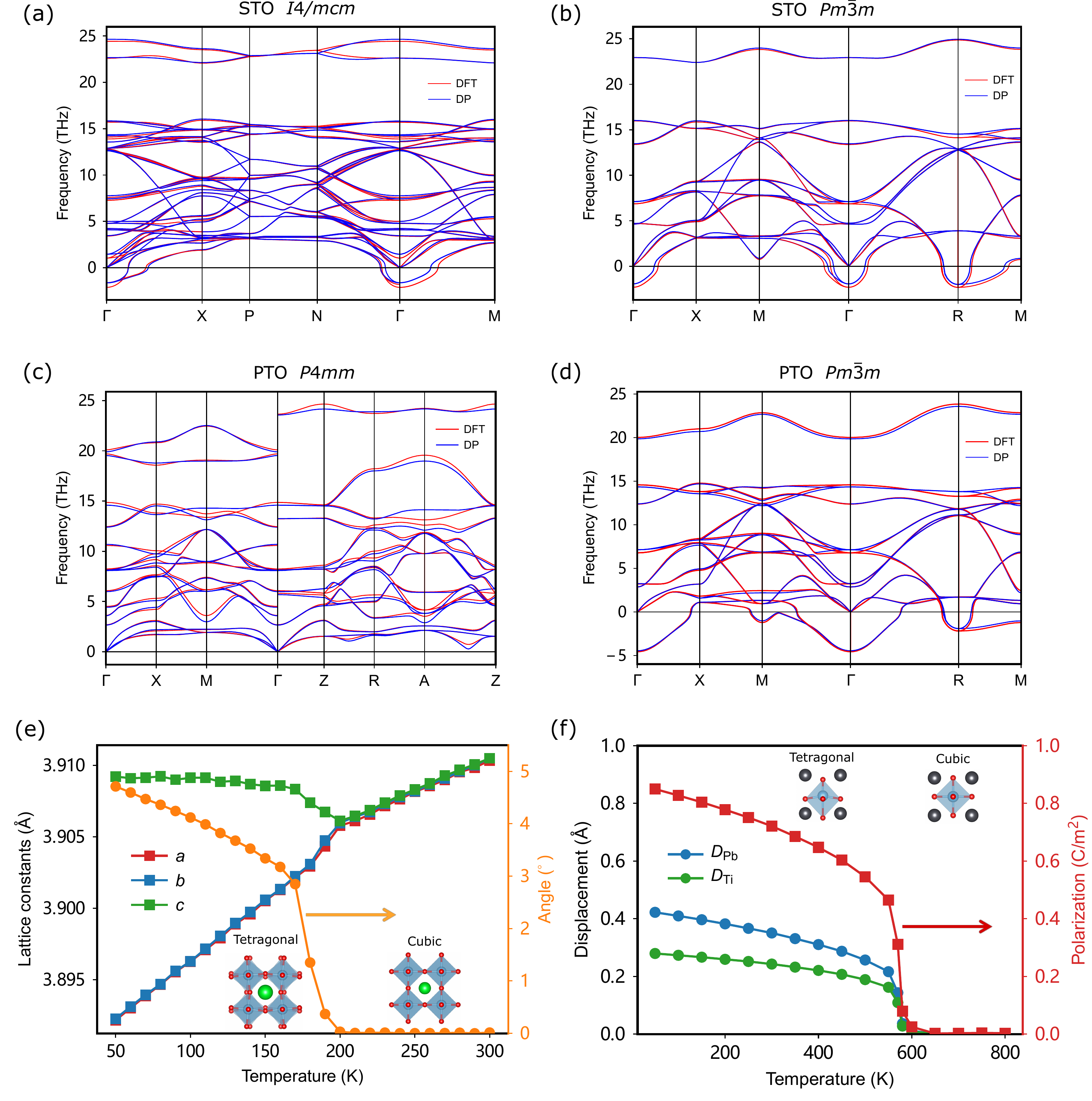}
\caption{Phonon spectra of (a) tetragonal STO, (b) cubic STO, (c) tetragonal PTO, and (d) cubic PTO. (e) Temperature dependence of lattice constants and TiO$_6$ tilt angle ($\phi$) in STO obtained with $NPT$ MD simulations using the DP model of~\PSTO. (f) Temperature dependence of spontaneous polarization and local atomic displacements of Pb and Ti ($D_{\rm Pb}$ and $D_{\rm Ti}$) in PTO from DPMD simulations.}
\label{Phonon_PSTO}
\end{figure}

\clearpage
\newpage
\begin{figure}[]
\centering
\includegraphics[width=0.8\textwidth]{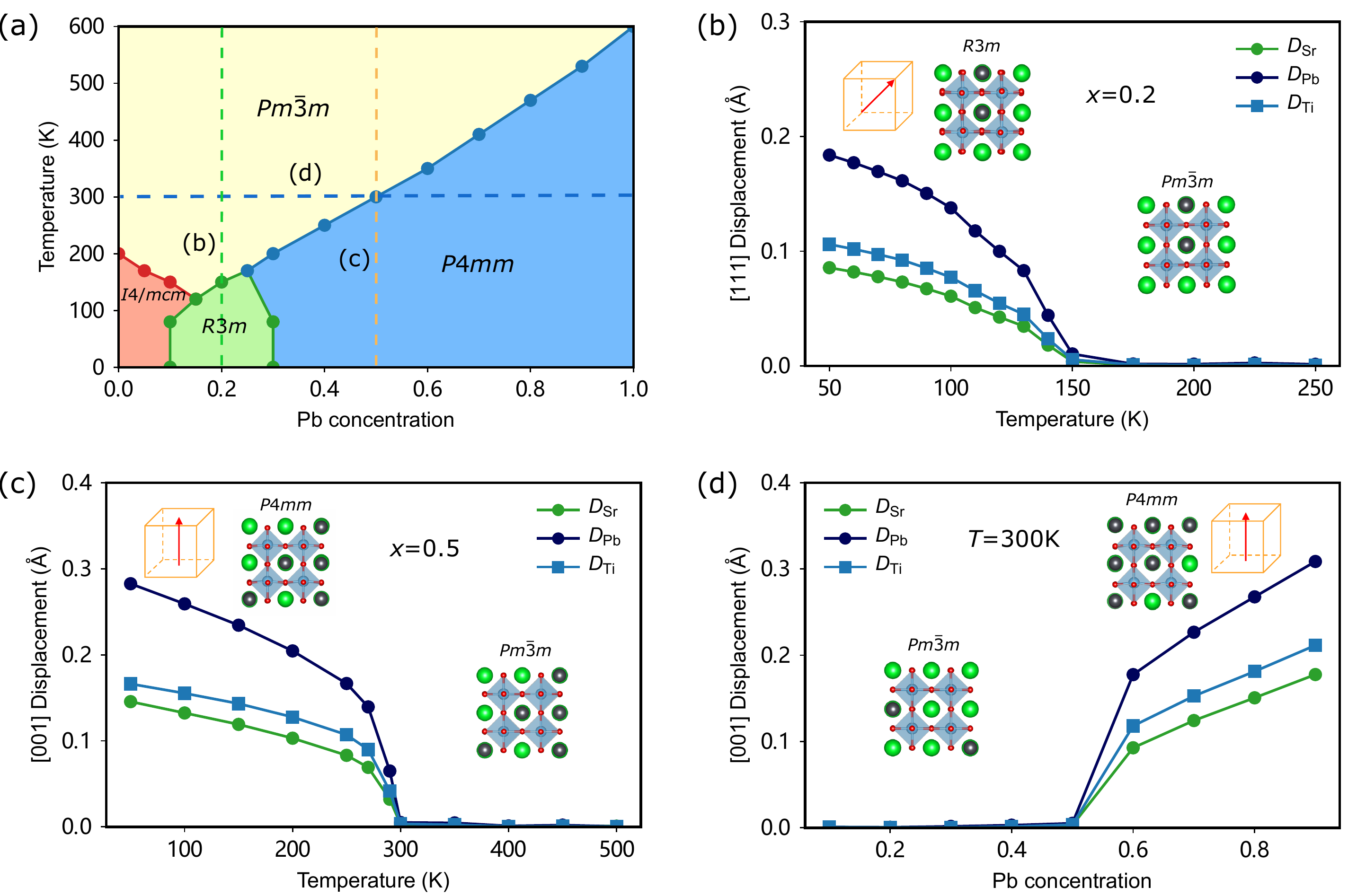}
\caption{(a) Temperature-composition phase diagram of \PSTO~solid solutions resulted from DPMD simulations. Temperature-driven phase transition in (b) Pb$_{0.2}$Sr$_{0.8}$TiO$_3$ and (c) Pb$_{0.5}$Sr$_{0.5}$TiO$_3$. (d) Composition-driven phase transition in \PSTO~at $T=300$~K.}
\label{PSTO_Tx}
\end{figure}

\clearpage
\newpage
\begin{figure}[]
\centering
\includegraphics[width=0.6\textwidth]{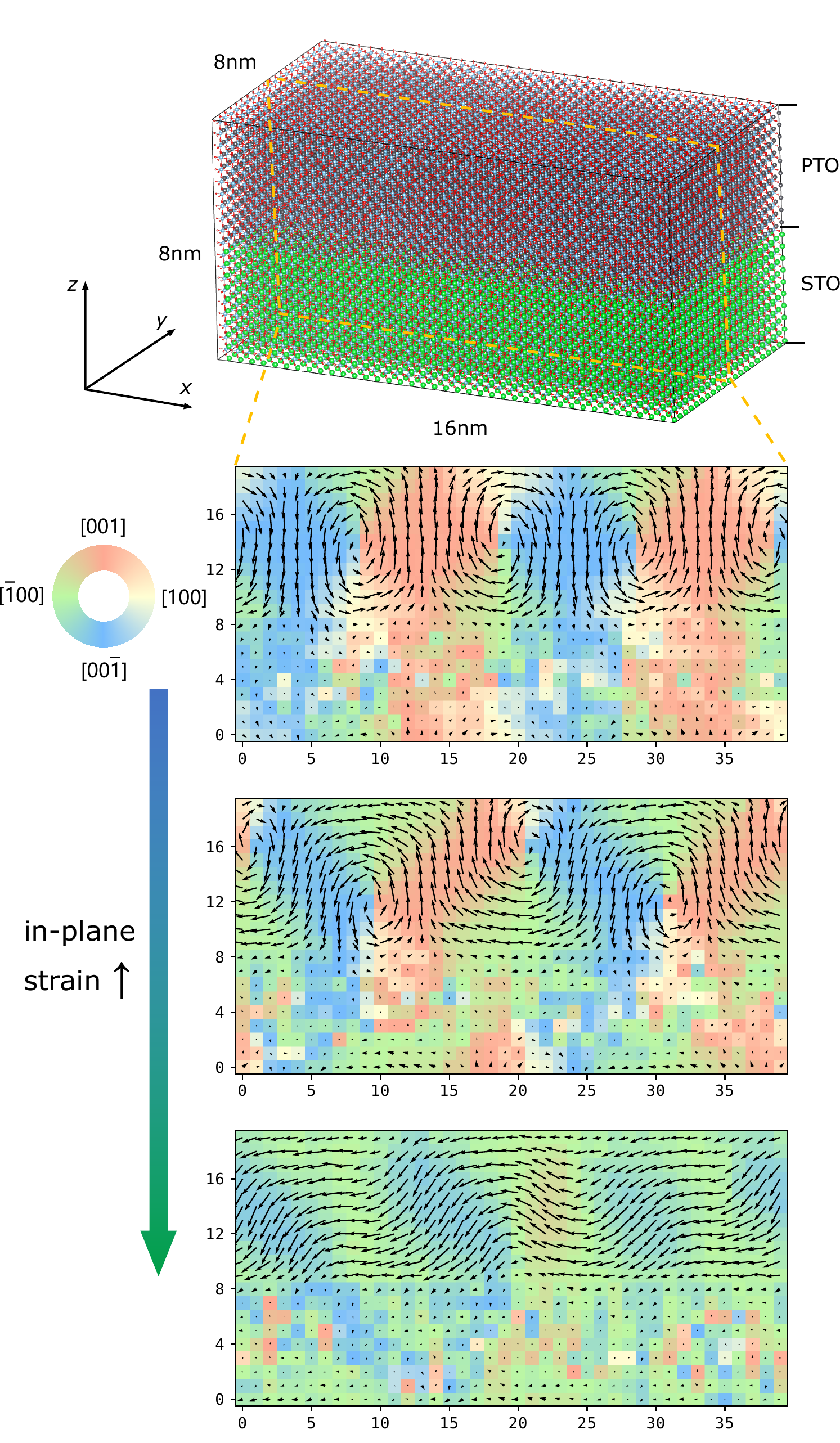}
\caption{(a) $40\times20\times20$ supercell of 80000 atoms with periodic boundary conditions used in DPMD to model (PTO)$_{10}$/(STO)$_{10}$ superlattices. The DP model of PSTO predicts a in-plane strain-driven transition from (b) ordered polar vortex lattice to (c) shifted polar vortex lattice, and to (d) electric dipole waves. Each black arrow represents the local electric dipole within a 5-atom unit cell and the background is colored based on the polarization direction.}
\label{Topo_struc}
\end{figure}

\clearpage
\newpage
\begin{figure}[]
\centering
\includegraphics[width=0.6\textwidth]{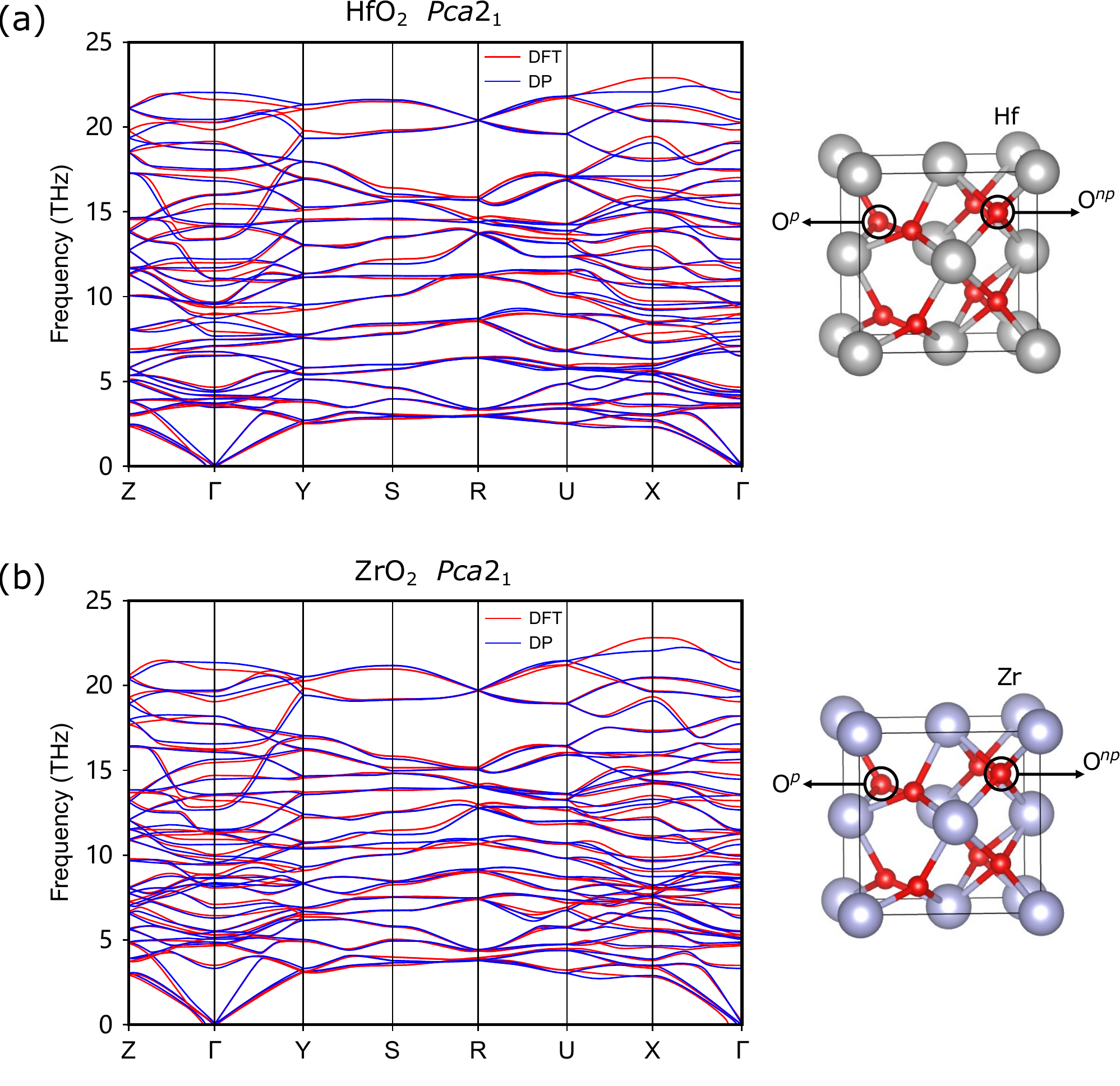}
\caption{Phonon spectra of (a) $Pca2_1$ HfO$_2$ and (b) $Pca2_1$ ZrO$_2$ computed with the DP model of HZO. The $Pca2_1$ phase has two types of oxygen atoms, three-fold coordinated polar oxygen (\OP) and fourfold-coordinated nonpolar oxygen (\ONP).
}
\label{Phonon_HZO}
\end{figure}

\clearpage
\newpage
\begin{figure}[]
\centering
\includegraphics[width=0.8\textwidth]{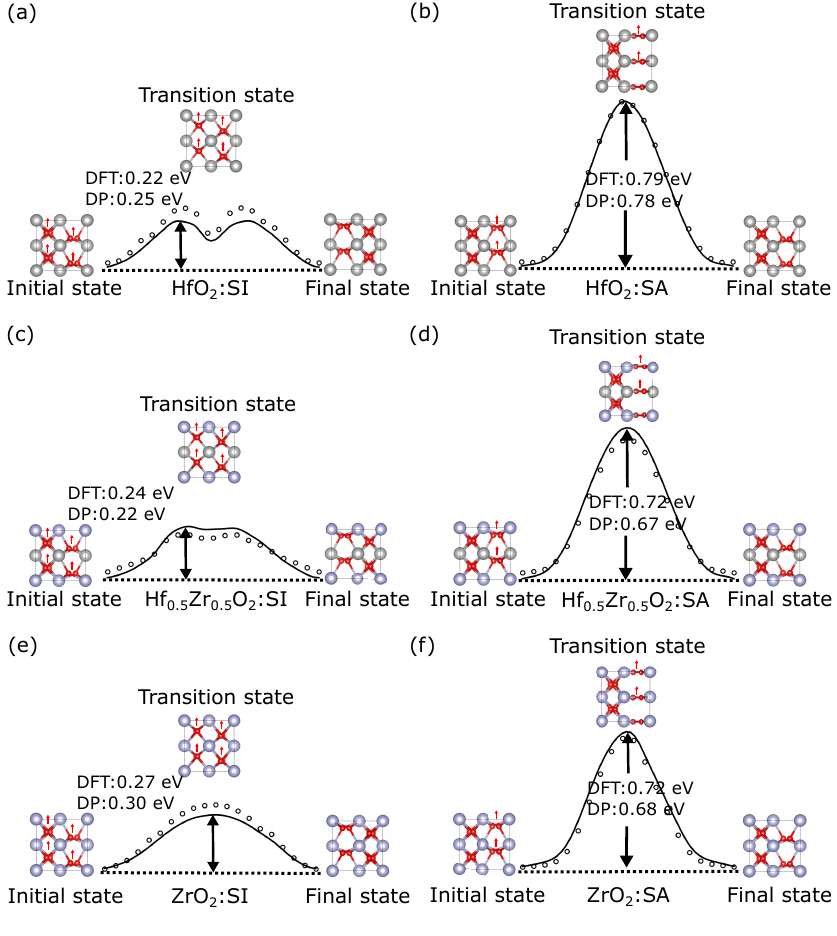}
\caption{ Comparison of the energy barriers of polarization reversal for shift-in (SI) and shift-across (SA) mechanisms in (a) HfO$_2$, (b) Hf$_{0.5}$Zr$_{0.5}$O$_2$, and (c) ZrO$_2$ calculated with DFT (solid line) and HZO DP model (empty circles).}
\label{HfO}
\end{figure}

\clearpage
\newpage
\begin{figure}[]
\centering
\includegraphics[width=0.8\textwidth]{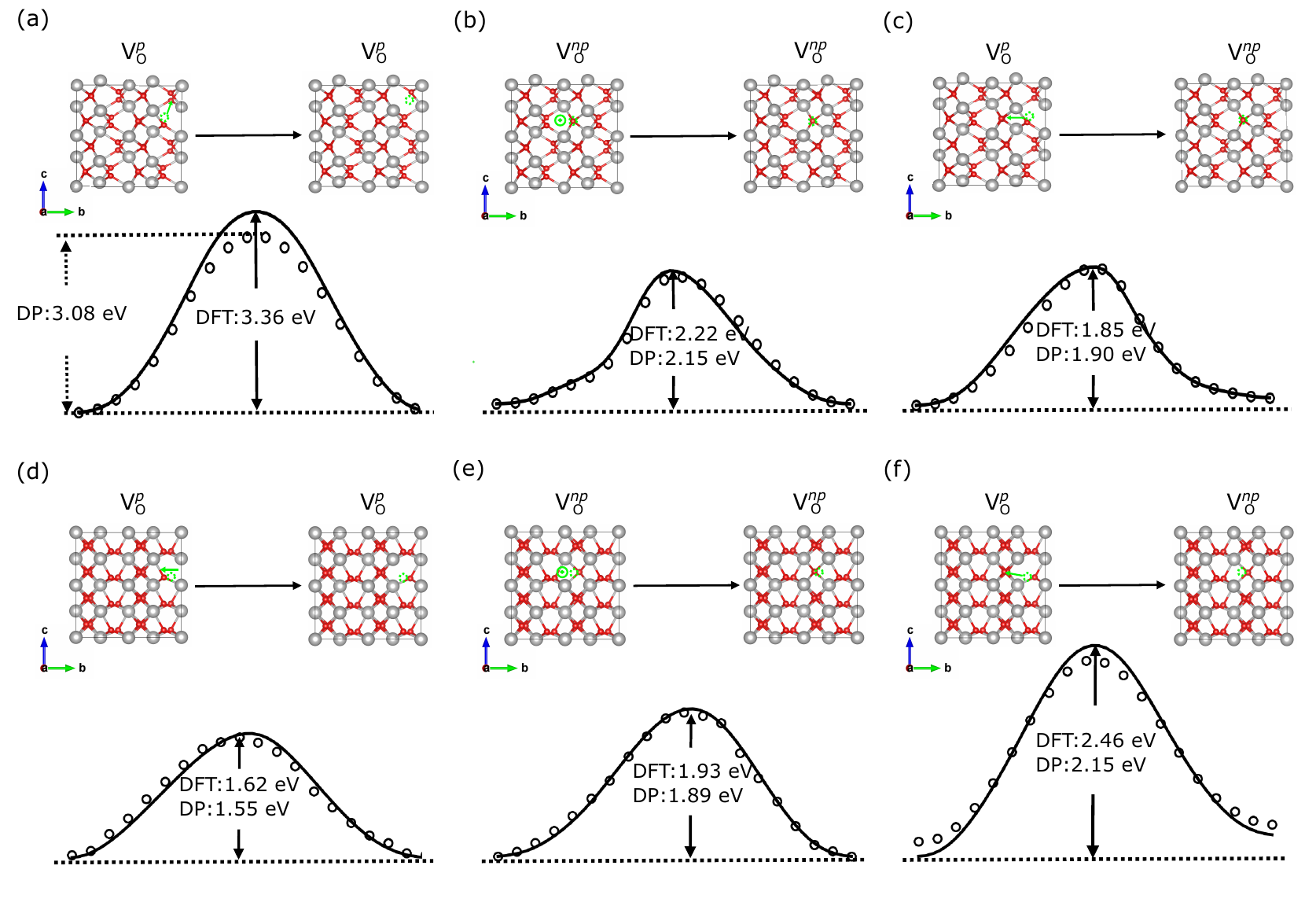}
\caption{Comparison of energy barriers of oxygen vacancy diffusion in (a-c) $P2_1/c$ and (d-f) $Pca2_1$ HfO$_{2-\delta}$ ($\delta=0.0625$) predicted by DFT (solid line) and DP (empty circles). The oxygen vacancy at the \OP~site is denoted as \VP and that at the \ONP~site is \VNP.}
\label{defect-NBE}
\end{figure}

\clearpage
\newpage
\begin{figure}[]
\centering
\includegraphics[width=0.8\textwidth]{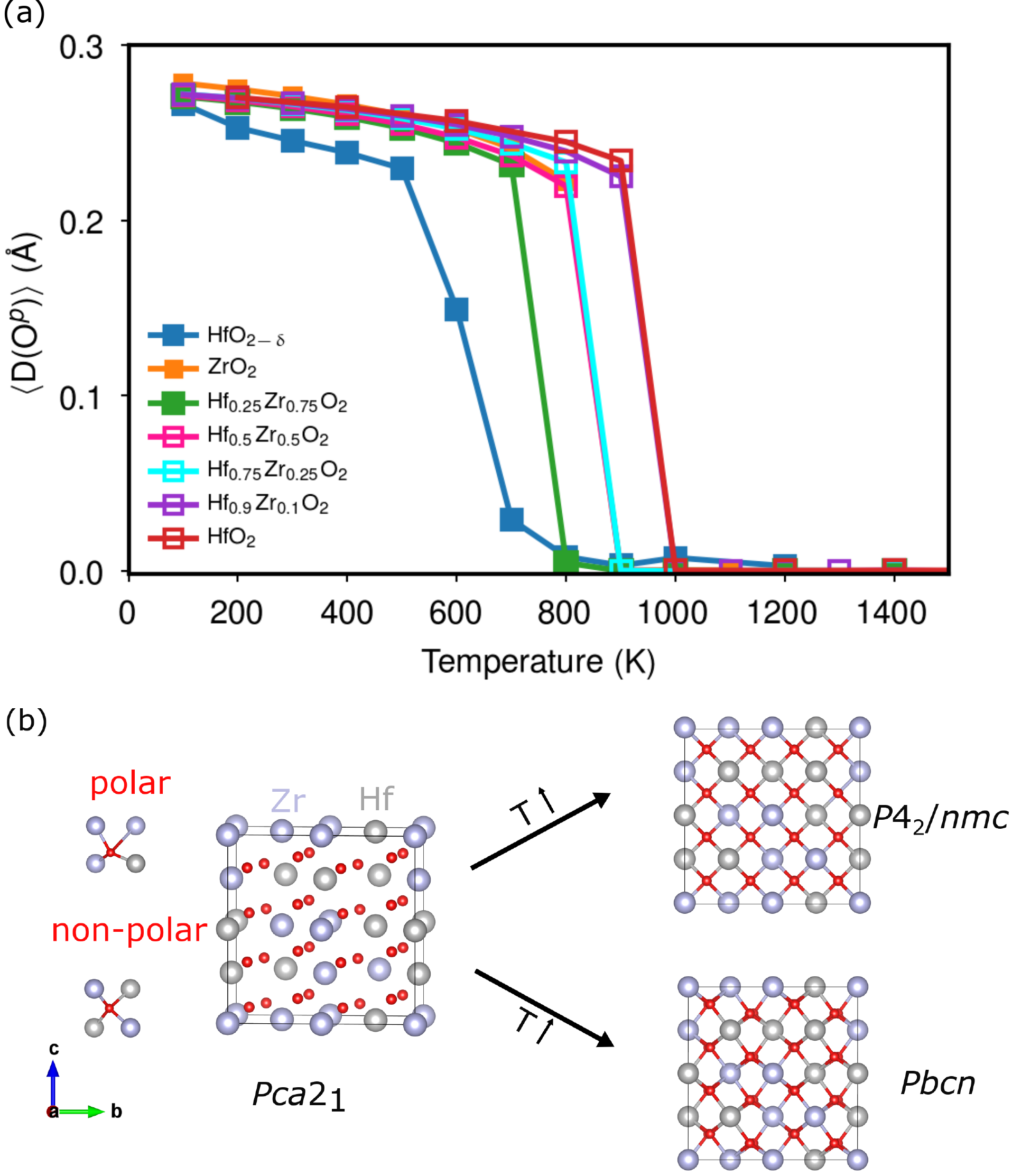}
\caption{(a) Temperature dependence of averaged polar oxygen displacement ($\left< D{({\rm O}^p)}\right>$) in HfO$_{2-\delta}$ ($\delta=0.0625$) and \HZO~($x=0.0, 0.25, 0.5, 0.75, 0.9, 1.0$) obtained from DPMD simulations of 6144-atom supercells using the same HZO model at a pressure of 1.0 bar. The starting configuration is the ferroelectric $Pca2_1$ phase. Solid markers denote $Pca2_1 \rightarrow P4_2/nmc$ while empty markers denote $Pca2_1 \rightarrow Pbcn$, both shown in (b).
}
\label{phase-transition}
\end{figure}
\end{document}